\documentclass[twocolumn,amsmath,amssymb,prd,reprint,longbibliography,floatfix,8pt]{revtex4-1}

\usepackage{cancel}
\usepackage{enumitem}
\usepackage[utf8x]{inputenc}
\usepackage{graphicx}
\usepackage{dcolumn}
\usepackage{bm}
\usepackage[switch]{lineno}
\usepackage{float} 
\usepackage{subfigure}
\usepackage{tikz}
\usepackage{multirow}
\usetikzlibrary{arrows.meta}
\usetikzlibrary{arrows,decorations.pathmorphing,backgrounds,positioning,fit,petri}
\usepackage{blochsphere}

\begin{document}

\title{Spin-phonon decoherence in solid-state paramagnetic defects from first principles}

\author{Sourav Mondal}
\author{Alessandro Lunghi}
\email{lunghia@tcd.ie}
\affiliation{School of Physics, AMBER and CRANN Institute, Trinity College, Dublin 2, Ireland}

\begin{abstract}
{\bf Paramagnetic defects in diamond and hexagonal boron nitride possess a unique combination of spin and optical properties that make them prototypical solid-state qubits. Despite the coherence of these spin qubits being critically limited by spin-phonon relaxation, a full understanding of this process is not yet available. Here we apply ab initio spin dynamics simulations to this problem and quantitatively reproduce the experimental temperature dependence of spin relaxation time and spin coherence time. We demonstrate that low-frequency two-phonon modulations of the zero-field splitting are responsible for spin relaxation and decoherence, and point to the nature of vibrations in 2-dimensional materials as the culprit for their shorter coherence time. These results provide a novel interpretation to spin-phonon decoherence in solid-state paramagnetic defects, offer a new strategy to correctly interpret experimental results, and pave the way for the accelerated design of new spin qubits.}
\end{abstract}

\maketitle

\section*{Introduction}

\begin{figure}[h!]
    \centering
    \includegraphics[scale=1.5]{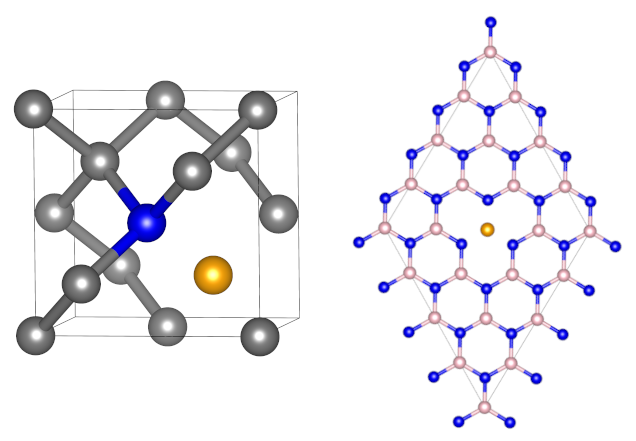}
    \caption{\textbf{Defects' Structure.} Left panel: The structure of the defects NV$^-$ in the unit cell of diamond. Right panel: the structure of V$_\mathbf{B}^-1$ embedded on a 6x6 supercell of h-BN. Carbon atoms are ported in grey, nitrogen atoms in blue, and boron atoms in pink. The vacancy is highlighted by the presence of a fictitious yellow atom.}
    \label{image0}
    \hfill
\end{figure}

Defects in solid-state semiconductors often introduce additional electronic states with energy lower than the band-gap, leading to color centers. Tens of different color centers are known for only diamond\cite{alkahtani2018fluorescent} and Silicon Carbide\cite{castelletto2020silicon} and their presence often enriches the original material's optical and magnetic properties, enabling interesting applications in the fields of sensing, photonics, and more. In particular, the negative nitrogen-vacancy (NV$^-$) center in diamond consists, represented in Fig. \ref{image0}, of a spin-1 system, with spin density localized at the defect site. NV$^-$ centers hold a special role in the quantum technology ecosystem thanks to some ideal properties, including long relaxation and coherence times ($T_1$ and $T_2$, respectively) at ambient temperature\cite{balasubramanian2009ultralong,bar2013solid}, chemical and mechanical robustness\cite{schirhagl2014nitrogen}, and the possibility to optically address and initialize the qubit states\cite{gruber1997scanning,jelezko2004observation}. NV$^-$ centers have thus found immediate application in quantum technologies, going from sensing\cite{grotz2011sensing,tetienne2013spin,schirhagl2014nitrogen} to quantum communications\cite{togan2010quantum}. On the verge of NV$^-$ success, other materials with similar properties have been individuated leading to a fast-growing family of solid-state spin qubits\cite{wolfowicz2021quantum}. This includes negative boron vacancies in hexagonal boron-nitride ($\mathrm{V_B^-}$), reported in Fig. \ref{image0}, which has emerged as a promising 2D-material-based qubit with an optimal spin-optical interface and mechanical properties\cite{gottscholl2020initialization,gottscholl2021room,stern2022room}.\\

Spin coherence is key for any quantum application and ultimately sets the limit of sensing accuracy or computation fidelity. Despite its central role in the physics of paramagnetic defects, the contribution of spin-phonon interaction to relaxation and decoherence has not yet been fully understood and experiments are invariably interpreted by means of phenomenological models based on a simplistic Debye picture of phonons\cite{walker19685,wolfowicz2021quantum}. Such a state of affairs effectively prevents the establishment of a rigorous understanding of spin dynamics in solid-state qubits and several outstanding questions are left unanswered. For instance, it is yet not clear what mechanism of spin relaxation is the driving one at non-cryogenic temperatures, with one- (Direct and Orbach) and two-phonon (Raman) mechanisms often interchangeably cited as possible relaxation routes\cite{jarmola2012temperature,gottscholl2021room}. Moreover, spin-phonon relaxation has been shown to limit spin coherence at ambient temperature, but with significant deviations from the commonly expected limit $T_2=2T_1$\cite{bar2013solid}, suggesting that additional vibronic contributions to decoherence might play an important role. Besides the importance of such questions from a fundamental science point of view, the current limits to our understanding of the microscopic processes determining spin coherence in paramagnetic solid-state materials pose serious challenges for their use as quantum sensors, namely one of the main applications for such compounds. In a typical experiment, the information on the spin's environment is obtained through the study of spin-environment fingerprints present in the spin dynamics\cite{degen2017quantum}. Such protocol is inherently connected to our ability to build reliable and quantitative models for spin-environment interactions and dynamics, of which spin-phonon coupling and relaxation are excellent examples. Last but not least, from a materials science point of view, the lack of robust structure-property relations for these compounds makes the identification of new paramagnetic defects with desired properties an outstanding challenge, so far mostly left to serendipity. \\

\begin{figure*}[t]
    \centering
    \includegraphics[scale=1]{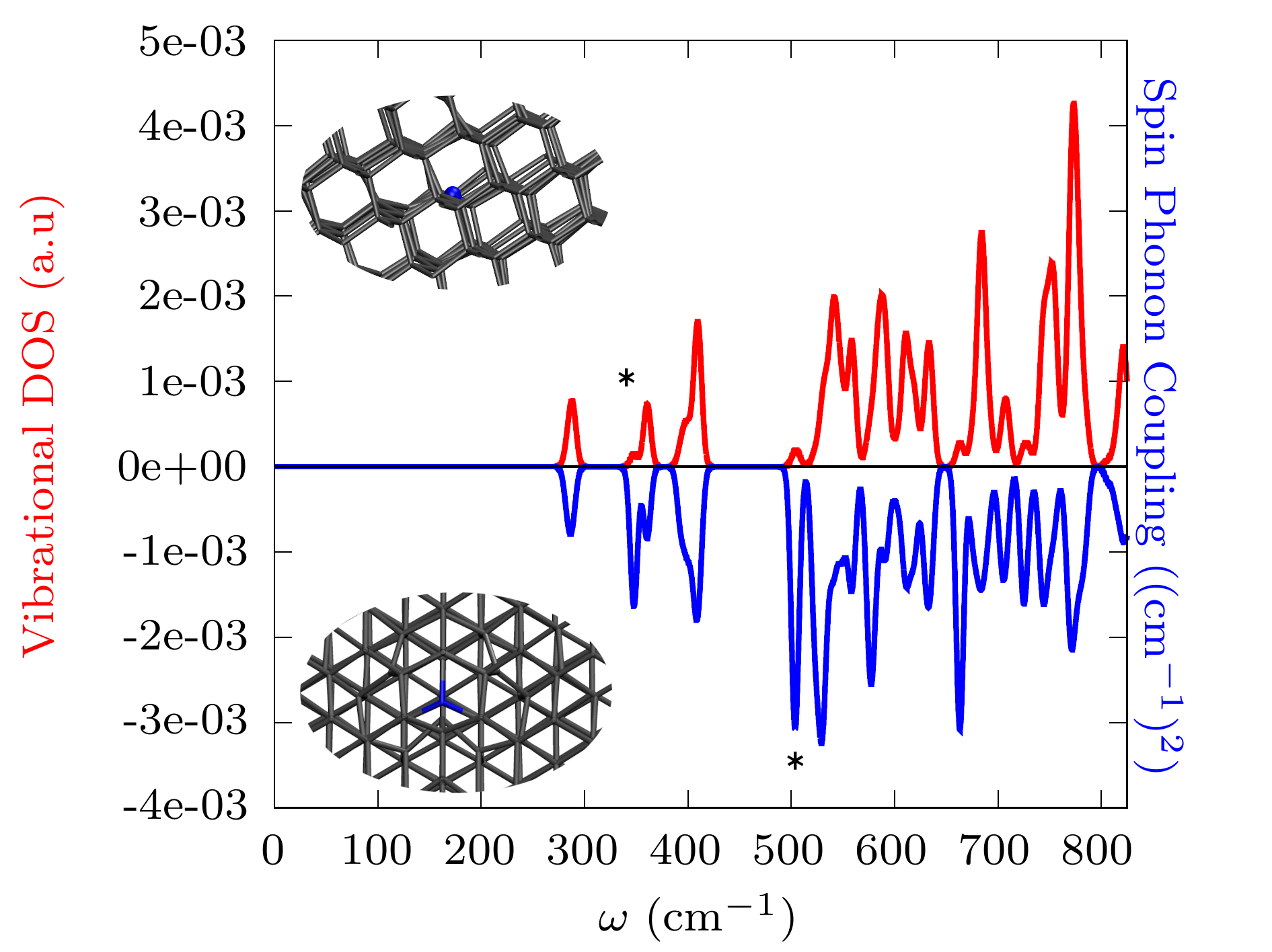}
    \includegraphics[scale=1]{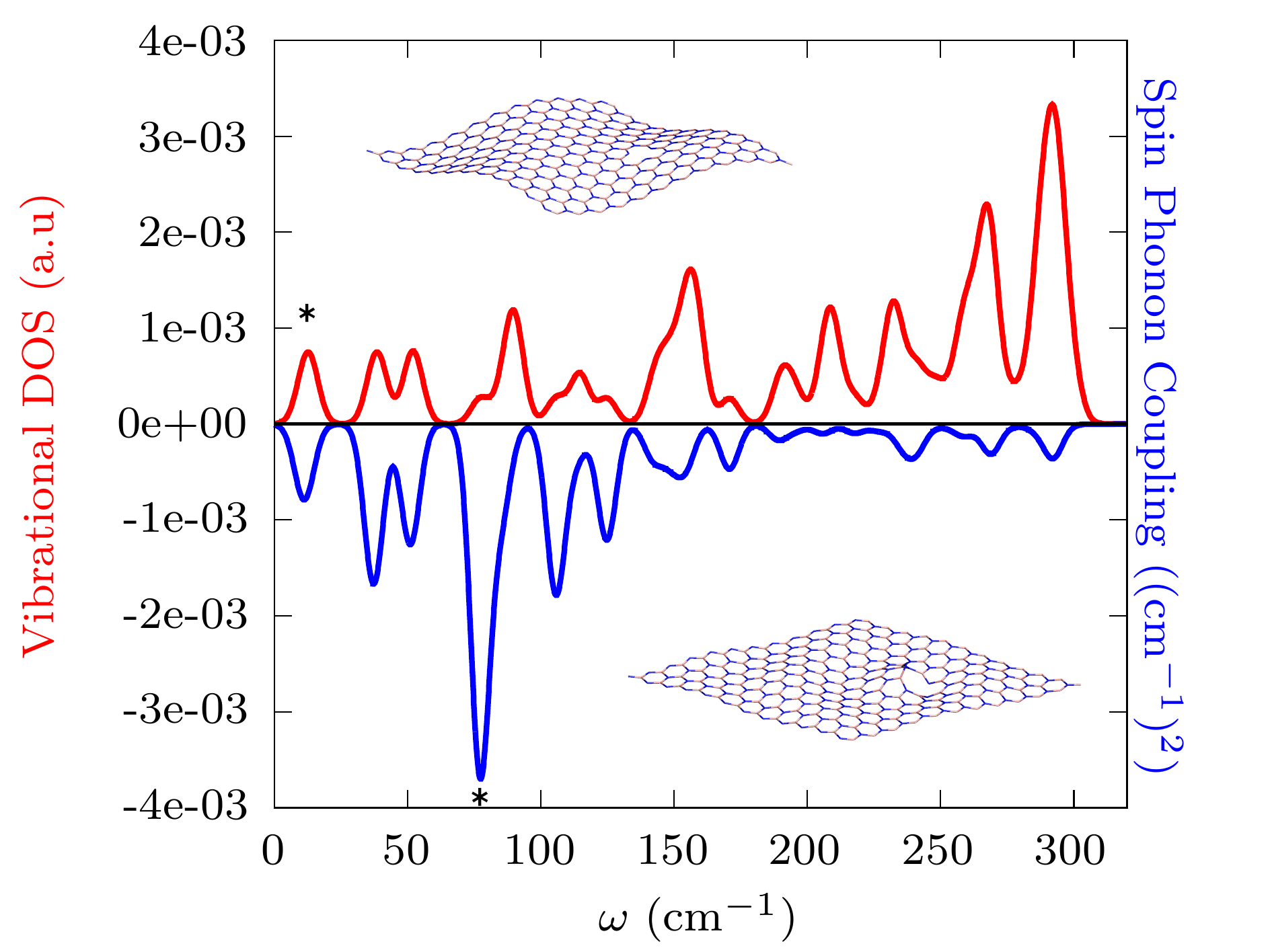}
    \caption{\textbf{Phonon density of state and spin-phonon coupling.} Results for NV$^-$ and V$_\mathbf{B}^-1$ are reported in the left and right panel, respectively. Red curves report the phonon density of states, while the blue curves report the spin-phonon coupling density $V(\omega)$. Star symbols point to the phonon modes displayed as insets.}
    \label{image1}
    \hfill
\end{figure*}

Ab initio methods have been applied to the problem of spin relaxation in the context of molecular systems in the last few years\cite{lunghi2019phonons,escalera2019exploring,lunghi2020limit,reta2021ab,lunghi2022toward,mondal2022unraveling}, leading to unprecedented insights into the physics of spin systems. On the ohter hand, only very recent attempts to apply ab initio strategies to solid-state defects are available\cite{astner2018solid,tang2022first,cambria2022temperature}, and a full picture of what is the leading relaxation mechanism at ambient temperature in both 3D and 2D semiconductors, the role of phonons on coherence, and what type of vibrations are responsible for such processes have not yet been fully elucidated. Here we extend the state-of-the-art in this nascent field\cite{lunghi2022spin,lunghi2022toward} to solid-state compounds and obtain a quantitative microscopic picture of spin relaxation and decoherence in NV$^-$ and $\mathrm{V_B^-}$. Our results can be exported to the rest of the increasingly large family of $S=1$ spin qubits\cite{wolfowicz2021quantum,bayliss2020optically} and have the potential to provide unprecedented guidance toward the design of new materials with optimal features for quantum technologies.

\section*{Results}

\textbf{Spin-Phonon Coupling.} The ground state of both NV$^-$ and V$_\mathrm{B}^-$ is well described by a $S=1$ Hamiltonian 
\begin{equation}
    \hat{H}_S= \vec{\mathbf{S}} \cdot \mathbf{D} \cdot \vec{\mathbf{S}} + \mu_\mathrm{B}\: \vec{\mathbf{S}} \cdot \mathbf{g} \cdot \vec{\mathbf{B}}
    \:,
    \label{spinHam}
\end{equation}
where the Bohr's magneton, $\mu_\mathrm{B}$, and the Lande's tensor, $\mathbf{g}$, mediate the interaction with the external field $\vec{\mathbf{B}}$, and $\vec{\mathbf{D}}$ is the zero-field splitting matrix. By expressing this Hamiltonian in a reference frame with the $z$-axis parallel to the defect's $C_3$ symmetry axis, the first term of Eq. \ref{spinHam} reduces to $\hat{H}_S= D S_z^2$, with $D \sim 0.118$ cm$^{-1}$ and $ \sim 0.092$ cm$^{-1}$ for V$_\mathrm{B}^-$, and NV$^-$, respectively, whereas $\mathbf{g}$ closely matches the free electron value for both compounds. The zero-field splitting stabilizes the $\hat{S}_z$ eigenstate $|0\rangle$ with respect to the pair $|\pm 1\rangle$ by an energy amount of $D$. Finally, the degeneracy of $|\pm 1\rangle$ is broken by the Zeeman interaction (see Fig. S6). Density functional theory (DFT) nicely reproduces these experimental values when applied to a small cluster of atoms around the defect (see SM). This approach is used to describe the defects' spin properties throughout this work. \\

The interaction between the defect's spin and the phonons is modeled by the spin-phonon Hamiltonian\cite{lunghi2022toward}
\begin{equation}
    \hat{H}_{\mathrm{S-Ph}}=\sum_{\alpha} \left( \frac{\partial \hat{H}_S}{ \partial q_{\alpha} } \right) q_{\alpha} + \frac{1}{2} \sum_{\alpha,\beta} \left( \frac{\partial^2 \hat{H}_S}{ \partial q_{\alpha} \partial q_{\beta}} \right) q_{\alpha} q_{\beta}\:,
    \label{SPHam}
\end{equation}
where we assume the phonons $q_{\alpha}$ to be quantum harmonic oscillators of frequency $\omega_{\alpha}/2\pi$. Here we consider the modulation of the zero-field splitting $\mathbf{D}$ as the source of spin-phonon coupling and numerically compute its first and second-order derivatives. In order to reduce the computational overheads associated with a second-order numerical differentiation, we trained a neural network on a set of $\mathbf{D}$ values computed with DFT over 1600 random distortions of the considered cluster\cite{lunghi2020limit,garlatti2022critical}. The neural network is tested over 400 residual distortions not used during the training and shows a remarkable ability to predict the values of $\mathbf{D}$ for unseen geometries with a root-mean-squared error of only 1.8 x 10$^{-4}$ cm$^{-1}$ for NV$^-$ and 2.6 x 10$^{-4}$ cm$^{-1}$ for V$_\mathrm{B}^-$, see Figs. S8-S10. The net is then used to perform a 6- and 36-point numerical differentiation of $\mathbf{D}$ for first- and second-order, respectively. \\

Phonons were computed with periodic DFT, and the vibrational density of states is shown in Fig. \ref{image1}. In order to individuate the coupling strength of each phonon we studied the function 
\begin{equation}
    V(\omega)= \sum_{ij}^{3} \left( \frac{\partial D_{ij}}{\partial q_{\alpha}} \right)^{2} \delta(\omega-\omega_\alpha)\:.
\end{equation}
$V(\omega)$ is plotted in Fig. \ref{image1}, together the phonon density of states, and shows that strongly coupled vibrations are present over the entire spectrum. A visual inspection of the strongest coupled mode for NV$^-$ shows that spin-phonon coupling is promoted by either compressive waves of diamonds (mode at $\sim$350 cm$^{-1}$ in Fig. \ref{image1}) and distortions localized at the defect site (mode at $\sim$500 cm$^{-1}$ in Fig. \ref{image1}). The analysis of modes for V$_\mathrm{B}^{-}$ shows that spin-phonon coupling is effectively originated by out-of-plane modes that locally distort the defect.\\

\textbf{Spin-Phonon Relaxation.} Spin-phonon relaxation is described by means of density-matrix perturbation theory, where phonons, namely the perturbation, act as a thermal reservoir with infinite specific heat and instantaneous relaxation time. In this framework, the Markovian time-evolution of the reduced spin density matrix in the interaction picture, $ \hat{\rho}_s(t)$, is described by\cite{lunghi2022toward}
\begin{equation}
    \frac{d\hat{\rho_s}}{dt} = \left( \hat{\hat{\mathbf{R2}}}^{1-\mathrm{ph}} + \hat{\hat{\mathbf{R2}}}^{2-\mathrm{ph}} + \hat{\hat{\mathbf{R4}}}^{2-\mathrm{ph}} \right) \: \hat{\rho}_s(t) \:,
\end{equation}
where $\hat{\hat{\mathbf{R2}}}^{1-\mathrm{ph}}$ describes resonant single-phonon processes arising from the first term of the Hamiltonian in Eq. \ref{SPHam} and second-order density matrix perturbation theory. As shown in Fig. \ref{image1}, phonons' energies largely exceed the zero-field splitting of the spin qubits, resulting in the absence of one-phonon processes, namely direct and Orbach relaxation mechanisms, which would require degeneracy among spin and phonon states. In principle, low-energy phonons would appear by increasing the size of the cell used for phonons simulations, thus recovering the long-wavelength acoustic modes of the lattice. However, as shown for molecular qubits\cite{lunghi2019phonons,garlatti2020unveiling}, these modes have vanishingly small spin-phonon coupling and become relevant to spin dynamics only in the low-temperature relaxation regime\cite{astner2018solid,gugler2018ab}. Low-temperature spin relaxation is also strongly affected by cross-relaxation, as shown by experiments conducted on samples with different defect concentrations\cite{jarmola2012temperature}, and does not represent an ideal regime to compare simulations and experiments. Therefore, we neglect the contribution $\hat{\hat{\mathbf{R2}}}^{1-ph}$ to spin dynamics and focus on the more application-relevant high-temperature limit dominated by two-phonon processes.\\

The term $\hat{\hat{\mathbf{R2}}}^{2-\mathrm{ph}}$ accounts for two-phonon processes leading to Raman relaxation and originates from 
the quadratic spin-phonon coupling terms of Eq. \ref{SPHam} at the second-order of perturbation theory. In this approximation, the dynamics of the diagonal elements of $\hat{\rho}_s$ reads 
\begin{equation}
R2^{2-\mathrm{ph}}_{bb,aa} = \frac{\pi}{4\hbar^{2}}  \sum_{\alpha\beta}  V^{\alpha\beta}_{ba}V^{\alpha\beta}_{ab}G^{2-\mathrm{ph}}(\omega_{ba},\omega_{\alpha},\omega_{\beta}) 
\label{Red22}
\end{equation}
where $V^{\alpha\beta}_{ab}$ stands for $\langle a | (\partial^{2} \hat{H}_{\mathrm{s}}/\partial q_{\alpha}\partial q_{\beta})  | b \rangle$. The function $G^{2-\mathrm{ph}}$ accounts for three possible processes involving two phonons: absorption of two phonons, emission of two phonons, and simultaneous emission of one phonon and absorption of a second one. Among all the possible two-phonon processes, two-phonon absorption/emission is negligible as they require phonons with energy lower than the zero-field splitting, absent in our framework. Therefore, the simultaneous emission/absorption of two phonons is the only process that enables energy-conserving two-phonon spin transitions, for which $G^{2-\mathrm{ph}}$ reads
\begin{equation}
    G^{2-\mathrm{ph}}(\omega_{ba},\omega_{\alpha},\omega_{\beta}) = \delta(\omega_{ba}-\omega_{\alpha}+\omega_{\beta})\bar{n}_{\alpha}(\bar{n}_{\beta}+1) \:. 
    \label{G2sph}
\end{equation}
Finally, the term $\hat{\hat{\mathbf{R4}}}^{2-\mathrm{ph}}$ accounts for those two-phonon processes due to linear spin-phonon coupling at the fourth-order perturbation theory and reads 
\begin{equation}
     R4_{bb,aa}^{2-ph}  =\frac{\pi}{2\hbar^2} \sum_{\alpha\beta}\left | T^{\alpha\beta,+}_{ba} + T^{\beta\alpha,-}_{ba} \right|^2G^{2-ph} (\omega_{ba}, \omega_{\alpha}, \omega_{\beta})\:,
    \label{Raman}
\end{equation}
where the terms
\begin{equation}
T^{\alpha\beta,\pm}_{ba} = \sum_{c} \frac{ \langle b| (\partial \hat{H}_{s}/\partial q_{{\alpha}}) |c\rangle \langle c| (\partial \hat{H}_{s}/\partial q_{{\beta}})|a\rangle }{E_c -E_a \pm \hbar\omega_\beta} \:,
\end{equation}
include the contributions of virtual excited states $|c \rangle$.\\

\begin{figure*}[t]
    \centering
    \includegraphics[scale=1]{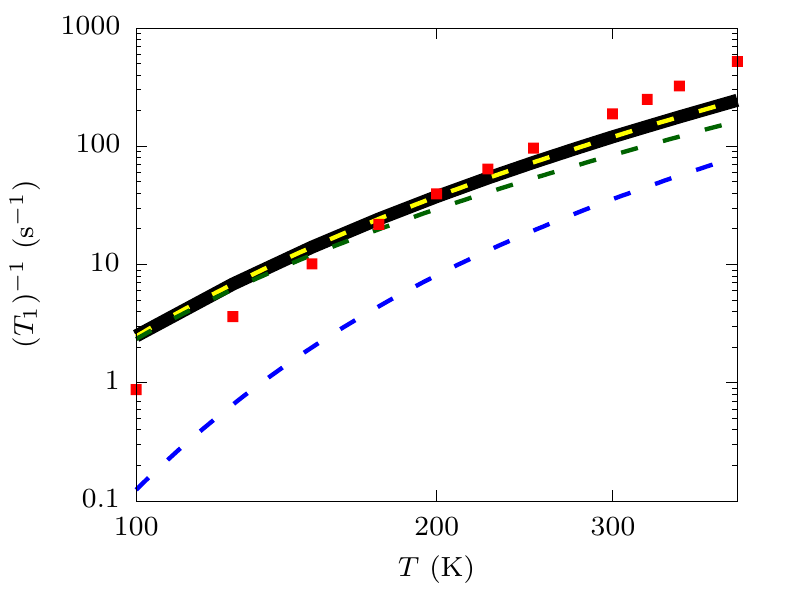}
    \includegraphics[scale=1]{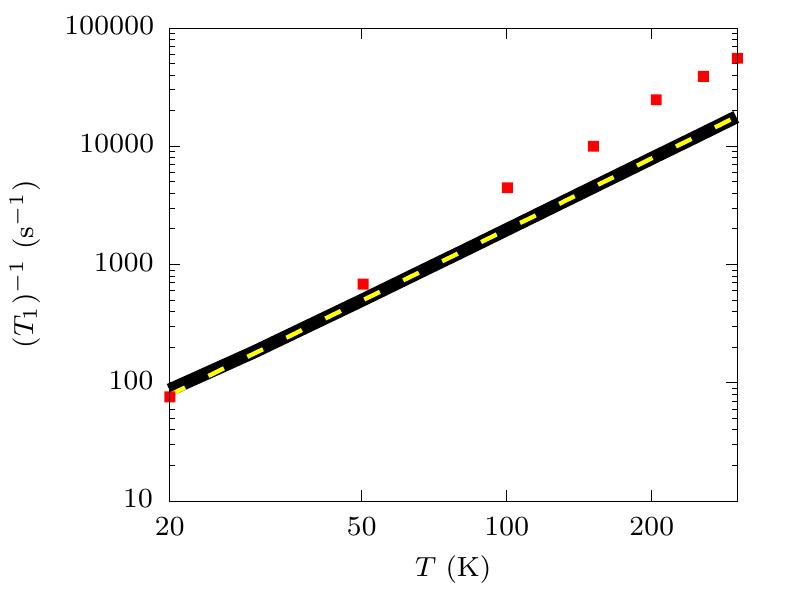}
    \caption{\textbf{ Spin-phonon relaxation.} 
    The left panel reports the simulated rate $(T_1)^{-1}$ (black solid line) as a function of $T$ for NV$^-$ defect in diamond together with the experimental rates (red squares)\cite{gottscholl2021room}. The dashed yellow line represents the fitting with Eq. \ref{fitTau}. Dashed blue and green lines are the two separate contributions of Eq. \ref{fitTau}. The right panel reports the simulated rate as a function of $T$ for V$_\mathrm{B}^-$ (solid black line) together with experimental rates (red squares). The dashed yellow line represents the fitting with the function $A\: T^{2}$.}
    \label{image2}
    \hfill
\end{figure*}

We compute $\hat{\hat{\mathbf{R2}}}^{2-\mathrm{ph}}$ and $\hat{\hat{\mathbf{R4}}}^{2-\mathrm{ph}}$ with the spin-phonon coupling coefficients and phonon frequencies determined by electronic structure calculations and simulate the relaxation dynamics of both defects with the software \textit{MolForge}\cite{lunghi2022toward,MolForge}. In accordance with the experimental setup, we initialize the system in $|1\rangle$ and then monitor the population recovery of $|0\rangle$ in time in the presence of a field of 1 mT\cite{jarmola2012temperature,gottscholl2021room}. As shown in Fig. S5 we observe a mono-exponential profile that leads to a single relaxation rate constant $(T_1)^{-1}$. Fig. \ref{image2} reports the simulated $(T_1)^{-1}$ as function of temperature obtained by applying $\hat{\hat{\mathbf{R2}}}^{2-\mathrm{ph}}$ to both defects. The comparison with experimental data reveals a striking agreement for both materials, and simulations nicely reproduce both the $T$-dependence and the trend in relaxation rates. On the other hand, simulations carried out with $\hat{\hat{\mathbf{R4}}}^{2-\mathrm{ph}}$ lead to much slower relaxation rates (see Fig. S4), thus leading to the conclusion that two-phonon relaxation due to quadratic spin-phonon coupling is the most relevant pathway in solid-state defects. The robustness of our conclusions is demonstrated by performing simulations with different phonons cell sizes, as shown in Fig. S2.\\

The excellent accuracy of simulations validates the computational approach and allows us to revisit the interpretation of spin relaxation in these systems. Firstly, our simulations reveal that no one-phonon Orbach process is involved in the spin relaxation of either NV$^-$ and V$_\mathrm{B}^-$ defects at high-temperature, as previously hypothesized in the fitting of experimental data\cite{jarmola2012temperature}. Our finding is in agreement with spin energy-conservation arguments and the low-energy cut-off frequency of the phonons' density of states. Moreover, the fitting of experimental data accounted for the Raman contribution to spin relaxation by means of a power-law expression $(T_1)^{-1} = A T^{5}$\cite{jarmola2012temperature}. Such expression was derived over 60 years ago by assuming the term $\hat{\hat{\mathbf{R4}}}^{2-\mathrm{ph}}$ as the dominating driving mechanism for Raman relaxation, a Debye-like structure of the phonons' density of states, and the absence of low-energy spin excited states\cite{walker19685}. Our simulations show that contrary to this model's assumptions, Raman relaxation does instead proceed through the term $\hat{\hat{\mathbf{R2}}}^{2-\mathrm{ph}}$ and that the phonons relevant to spin relaxation are not at all well described by a Debye model. The analysis of our simulations indeed shows that a strict $(T_1)^{-1} \propto T^{5}$ relation does not hold and that the $T$-dependence of spin relaxation has a more complex profile. As suggested by the $T$-dependence of the function $G^{2-\mathrm{ph}}$, we successfully modelled the simulations of NV$^{-}$ by means of the expression
\begin{equation}
    (T_1)^{-1} = A \frac{ \: e^{-B/k_{B}T}}{(e^{-B/k_{B}T}-1)^2} + C\frac{ \: e^{-D/k_{B}T}}{(e^{-D/k_{B}T}-1)^2}\:,
    \label{fitTau}
\end{equation}
where the fitted coefficients $C=326$ cm$^{-1}$ and $D=576$ cm$^{-1}$ nicely correspond to the first few peaks in the spin-phonon coupling density $V(\omega)$ reported in Fig. \ref{image1}, in qualitative agreement with a recent work\cite{cambria2022temperature}. The same type of expression would apply to V$_{B}^{-}$, but differently from NV$^{-}$, the spin relaxation of the V$_{B}^{-}$ is determined in the high-$T$ regime ($k_{B}T > \hbar\omega$) with respect to the first available modes, thus leading to a simple $(T_1)^{-1}=A\: T^{2}$ contribution, also in nice agreement with the experimentally determined $T^{2.5}$ power-law\cite{gottscholl2021room}. To corroborate the interpretation that the first available phonons are indeed the prominent ones for spin relaxation, we simulate $T_1$ by imposing a variable cut-off on the energy of phonons included in the computation of $\hat{\hat{\mathbf{R2}}}^{2-\mathrm{ph}}$ and show that even the inclusion of the first few available phonons is enough to obtain the correct order of magnitude of relaxation rate, with the inclusion of phonons up to 800 cm$^{-1}$ and 80 cm$^{-1}$ to entirely recover the converged value of $T_1$ for NV$^-$ and $V_\mathrm{B}^-$, respectively (see Fig. S7). This is in agreement with previous findings for magnetic molecules\cite{lunghi2022toward} and provides a simple rationale for explaining the striking difference among the relaxation rate of NV$^-$ and V$_\mathrm{B}^-$. Indeed, despite being both materials with stiff chemical bonds, the 2D nature of h-BN enables low-energy out-of-plane flexural vibrations that are sensibly populated even at low-$T$. \\

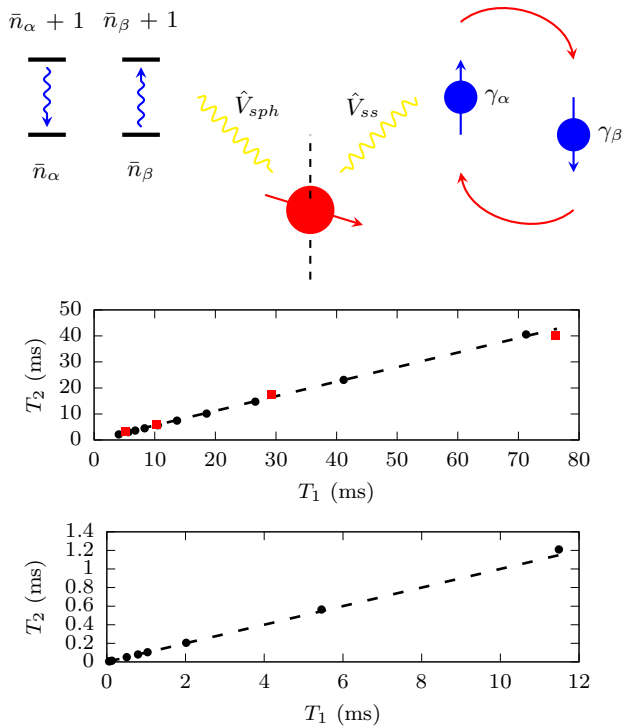
\begin{figure}[h!]
    \centering

\begin{tikzpicture}

\draw [ultra thick] (-4.75,2) -- (-4.25,2);
\draw [ultra thick] (-4.75,3) -- (-4.25,3);
\draw [thick,blue,->,>=stealth,decorate, decoration={snake,amplitude=.4mm,segment length=2mm,post length=1mm}] (-4.5,2.9) -- (-4.5,2.1) ;  
\node at (-4.5,1.5) {$\bar{n}_{\alpha}$};
\node at (-4.5,3.5) {$\bar{n}_{\alpha}$ + 1 };

\draw [thick,blue,->,>=stealth,decorate, decoration={snake,amplitude=.4mm,segment length=2mm,post length=1mm}] (-3.25,2.1) -- (-3.25,2.9) ;  
\draw [ultra thick] (-3,2) -- (-3.5,2);
\draw [ultra thick] (-3,3) -- (-3.5,3);
\node at (-3.25,1.5) {$\bar{n}_{\beta}$};
\node at (-3.25,3.5) {$\bar{n}_{\beta}$ + 1 };

\draw [thick,yellow,>=stealth,decorate, decoration={snake,amplitude=.8mm,segment length=2mm}] (-2.5,2.5) -- (-1.5,1.5) ;  
\node at (-1.7,2.4) {$\hat{V}_{sph}$};
\node at (-0.3,2.4) {$\hat{V}_{ss}$};

\draw [thick,yellow,>=stealth,decorate, decoration={snake,amplitude=.8mm,segment length=2mm}] (0.4,2.5) -- (-0.6,1.5) ;

\filldraw[color=red, fill=red, very thick](-1,1) circle (0.3);
\draw [thick,red,->,>=stealth] (-1.6,1.2) -- (-0.3,0.8) ;
\draw [thick,black,dashed] (-1,1.15) -- (-1,2)  ;
\draw [thick,black,dashed] (-1,0.6) -- (-1,0) ;

\draw [thick,red,->,>=stealth] (1,3.5) to [out=40,in=100] (2.5,3) ; \draw [thick,red,->,>=stealth] (2.5,1) to [out=220,in=280] (1,1.5) ;
\draw [thick,blue,->,>=stealth] (1,2) -- (1,3);  
\draw [thick,blue,<-,>=stealth] (2.5,1.5) -- (2.5,2.5);  
\filldraw[color=blue, fill=blue, very thick](1,2.5) circle (0.2);
\filldraw[color=blue, fill=blue, very thick](2.5,2) circle (0.2);
\node at (1.5,2.5) {$\gamma_{\alpha}$};
\node at (3,2) {$\gamma_{\beta}$};

\end{tikzpicture}
    
    \includegraphics[scale=1]{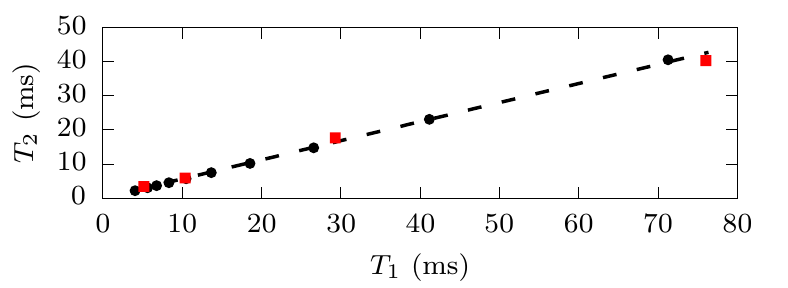}\\
    \includegraphics[scale=1]{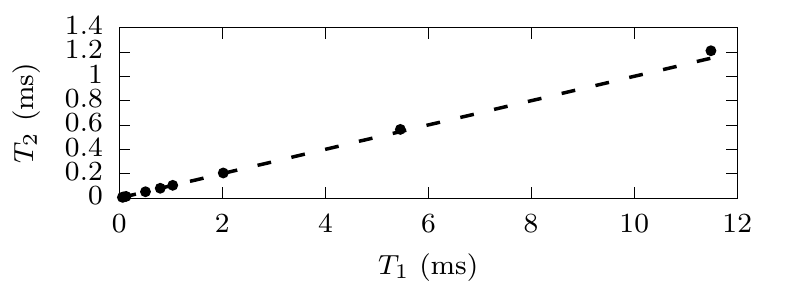}
    \caption{\textbf{Pure dephasing mechanism and correlations between spin-phonon relaxation time (T$_1$) and coherence time (T$_2$).} Top panel: schematic representation of two-phonon pure dephasing mechanism. Two degenerate phonons exchange population and lead to dephasing through the spin-phonon coupling interaction $\hat{V}_{sph}$, similarly to the pure dephasing due to nuclear spin flip-flop processes among spins with identical gyromagnetic factors $\gamma_\alpha=\gamma_\beta$. Middle panel: Black and red symbols represent the simulated and experimental\cite{bar2013solid} results for NV$^-$. The dashed line reports $T_2= 0.56 T_1$. Bottom panel: Black symbols represent the simulated results for V$_\mathrm{B}^-$. The dashed line reports $T_2= 0.1 T_1$.}
    \label{image3}
    \hfill
\end{figure}

\textbf{Spin-Phonon Decoherence.} Finally, we turn to the study of the limits imposed to spin coherence by spin-phonon relaxation by studying the dynamics of the out-of-diagonal elements of $\hat{\rho}_s$. Considering the effect of the super-operator $\hat{\hat{\mathbf{R2}}}^{2-\mathrm{ph}}$ we obtain the analytical result
\begin{equation}
 (T_2)^{-1}=(2T_1)^{-1} + (T_2^*)^{-1}\:,
\end{equation}
where the first term on the r.h.s. is the commonly reported upper bound of coherence due to a finite spin lifetime. The term $T_2^*$ represents the pure-dephasing term and reads
\begin{align}
    \frac{1}{T_2^*} \propto & \sum_{\alpha,\beta} \left( -2V^{\alpha\beta}_{aa}   V^{\alpha\beta}_{bb}+V^{\alpha\beta}_{aa} V^{\alpha,\beta}_{aa} + V^{\alpha\beta}_{bb} V^{\alpha\beta}_{bb} \right) \cdot \\ & \cdot G^{2-\mathrm{ph}}(0,\omega_\alpha,\omega_\beta)\:,
    \label{dephas}
\end{align}
where $a,b=\{ | 0 \rangle, | 1 \rangle \}$. Such a mechanism thus involves an energy-conserving simultaneous emission and absorption of two degenerate phonons. If the two phonons have different spin-phonon coupling, this process leads to decoherence, equivalently to the flip-flop nuclear spin transitions limiting the coherence of a central electronic spin\cite{yao2006theory,witzel2006quantum,seo2016quantum}. These two processes are schematically shown in the top panel of Fig. \ref{image3}
Notably, the corresponding expression of pure dephasing for one-phonon processes leads to a null contribution, as energy conservation would require zero-frequency phonons. Recently, it has been shown that pure spin dephasing is an active decoherence mechanism in molecular $S=1/2$ spin qubits\cite{garlatti2022critical}. Here we test the possibility for this process to also be influencing the values of $T_2$ for NV$^{-}$ observed in dynamical decoupling experiments\cite{bar2013solid}. In such experimental conditions, the contribution of spin-spin dipolar interactions to dephasing is systematically removed by applying several refocusing $\pi$ pulses, revealing the contribution of phonons to decoherence and allowing us to compare simulations and experimental results. We study the evolution of the state $|\Psi\rangle=( |0\rangle + |1\rangle )/\sqrt{2}$, as generated in experiments through a $\pi/2$ pulse resonant with the transition $|0\rangle \rightarrow |1\rangle$\cite{bar2013solid}. The middle panel of Fig. \ref{image3} reports the ratio between the values of $T_2$ and $T_1$ at different values of temperature for NV$^-$. The comparison between simulations and experimental results shows a remarkable agreement between the two, with simulations correctly capturing the experimentally observed trend $T_2 \sim 0.5T_1$, demonstrating the origin of such relation in the presence of pure dephasing. Interestingly, the simulations for V$_\mathrm{B}^-$ reported in the bottom panel of Fig. \ref{image3} show an even more drastic reduction of $T_2$, namely $T_2\sim 0.1 T_1$.

\section*{Discussion}

The theory of spin relaxation in solid-state paramagnetic defects or impurities has a long-standing tradition with roots in the early days of magnetic resonance. Seminal contributions to the theory of spin relaxation\cite{van1940paramagnetic} were rapidly followed by a proliferation of phenomenological models based on varied assumptions about the details of spin-phonon coupling and materials' vibrational density of states\cite{shrivastava1983theory,lunghi2022spin}. Until very recently, such models have remained the only tool available for the interpretation of experimental relaxation data. However, experimental data has proved to be far from easy to interpret and multiple models can often be equally good explanations of experimental evidence. The field of paramagnetic defects is no exception and no clear understanding of spin relaxation in this class of materials has yet been established. Providing a clear interpretation of the mechanism of spin relaxation is one of the main results of this work. Ab initio simulations, free from any adjustable parameter, enabled the identification of two-phonon Raman relaxation due to the quadratic coupling of low-energy optical phonons and zero-field splitting as the main relaxation pathway for NV$^-$ and V$_\mathrm{B}^-$. Although alternative relaxation pathways might be operative in spin qubits with different spin multiplicities or low-lying excited electronic states\cite{lunghi2022toward}, these results are likely exportable to all spin qubits exhibiting zero-field splitting, as suggested by their validity for defects in both 3D and 2D materials. As a consequence of the correct reinterpretation of the spin relaxation mechanism, we were also able to unravel the role of phonons on spin decoherence. The latter is invariably found to be bound from above by $T_1$ \cite{bar2013solid,simin2017locking} instead of the way around, as suggested by one-phonon relaxation theories. As shown, two-phonon processes introduce a pure dephasing mechanism and naturally explain such apparent deviations, leading to a coherent picture of spin dynamics. \\

The present results have far-reaching consequences and suggest several possible pathways for future development. For instance, the relevance of low-energy phonons for the coherence dynamics of $S=1$ spin qubits has several important consequences for the design of novel spin qubits. Although bulk diamond probably already offers the best possible environment possible in terms of the (high) energy of optical vibrations, new spin qubits can be searched among materials with high-energy optical vibrations and a natural low density of nuclear spins\cite{kanai2022generalized} in order to simultaneously minimize the effect of both spin-phonon coupling and spin-spin dipolar interactions.

Moreover, our results suggest that low-energy vibrational states of shallow defects and nano-diamonds might play a role in reducing the coherence time of electron spins in these substrates \cite{naydenov2011dynamical}. Indeed, it has recently been shown that the drastically reduced $T_1$ observed for NV$^-$ nano-diamonds has a temperature profile consistent with phonon-induced relaxation\cite{de2020temperature}. DeGuillebon et al. invoked the presence of magnetic impurities at the surface of the nano-diamonds to interpret their results\cite{de2020temperature}, but our simulations suggest that surface vibrational states, qualitatively similar to those of V$_\mathrm{B}^-$, might be enough to explain the drastic reduction in $T_1$. 2D materials hetero-structures and tailored surface coating might also offer a rich playground to tune the vibrational properties of spin qubits' host environment. 

Overall, our simulations open a window on the fine details of spin-phonon coupling in solid-state paramagnetic defects that can eventually inform several applications. Indeed, the coupling of vibrations and electronic states are responsible for the temperature-dependent shift in spin energy transitions often used as a sensing strategy of the spin environment\cite{acosta2010temperature}, as well as for the optical polarization of defects' spin states\cite{thiering2018theory}. Moreover, whilst minimizing the effect of spin-phonon coupling is a key future challenge to increase coherence time at high-temperature, a tailored strong coupling with surface or bulk acoustic waves holds great promise towards the control and coupling of multiple spin systems in solid-state architectures\cite{albrecht2013coupling,golter2016optomechanical,whiteley2019spin}. The proposed computational framework has a strong potential to aid the development of all these areas.\\

In conclusion, we have here used ab initio spin dynamics simulations to determine the spin-phonon coupling coefficients in two prototypical solid-state qubits and demonstrated how this interaction leads to spin relaxation and decoherence. By using a fully parameter-free computational approach we have provided novel insights into the nature of two-phonon Raman relaxation and revealed low-energy phonons as the main source of decoherence in solid-state qubits at high temperatures. We anticipate that our results will facilitate the interpretation of experimental relaxation data in spin qubits and speed up the identification of novel quantum sensors.

\section*{Materials and Methods}

{\bf Crystal Structures.} The defects' structure has been generated starting from a supercell of the pristine materials, diamond and h-BN, respectively. Cell optimizations and simulation of phonons for both the defects have been performed for two sizes of the supercell, i.e 3x3x3 and 4x4x4 diamond supercells for NV$^-$ and of 9x9 and 12x12 h-BN supercells for V$_\mathrm{B}^-$. Supercells of 3x3x3 and 4x4x4 for NV$^-$ contain 215 atoms and 511 atoms, respectively, whereas, 9x9 and 12x12 supercells for  V$_\mathrm{B}^-$ contain 161 atoms and 287 atoms. The supercells are shown in Fig. S1.\\

{\bf Electronic structure simulations.} Cell optimizations and phonon calculations for all the supercells of NV$^-$ defect in diamond and V$_\mathrm{B}^-$  defect in h-BN were performed within the framework of periodic density functional theory using CP2K.\cite{kuhne2020cp2k} All the optimization and phonon calculations were done at $\Gamma$-point.  A very tight force convergence criteria of 10$^{−7}$ a.u. and SCF convergence criteria of 10$^{-10}$ a.u. for the energy were employed for the cell optimizations. A plane wave cutoff of 1000 Ry with DZVP-MOLOPT Gaussian basis sets and Goedecker-Tetter Hutter pseudopotentials\cite{goedecker1996separable} were used for both systems. The Perdew-Burke-Ernzerhof (PBE) functional and DFT-D3 methods for dispersion corrections had been used.\cite{perdew1996generalized,grimme2010consistent}\\

Density Functional Theory, as implemented in the software ORCA\cite{neese2020orca}, had been used to compute the anisotropy tensor $\mathbf{D}$.  Clusters (passivated with H) created from atoms around the defect had been used for this purpose. A convergence test for different sizes of clusters had been performed to select the smallest cluster size which can reproduce the experimental $D$ value, see Table S1.
A similar strategy had been taken for the choice of the basis set and the DFT functional. Convergence tests had been performed to select the best ones in comparison to the experiments. A detailed comparison of different basis sets and different functionals has been presented in the table below, see Table. S2. Based upon the comparison, we have used 5 \AA$ $ cluster and  BHANDHLY \cite{becke1993new} functional for NV$^-$ defect and 7.5 \AA$ $ cluster and PBE0 \cite{adamo1999toward} functional for V$_\mathrm{B}^-$  defect, respectively. The basis set of DKH-def2-SVP had been used for all the atoms present in both the defects NV$^-$ and V$_\mathrm{B}^-$. The hydrogenated clusters  of NV$^-$ defect (5 \AA$ $) and V$_\mathrm{B}^-$ (7.5 \AA$ $) defect have been shown in Fig. S1.\\\\

{\bf Training of Neural Network and Test set generation.}
The training sets for the neural network have been generated by calculating $D$ for over 2000 distorted structures. The distorted structures were obtained by displacing all the Cartesian coordinates of all the atoms from the optimized geometry. The displacement was given by a random value in the range from $-$0.05 \AA$ $ to 0.05 \AA$ $. Among the total of 2000 distorted geometries, 1600 were used for training the ML models, while 400 were used for validation. The neural network we used  has 3 hidden layers consisting of 128, 64, and 16 nodes, respectively. The input layer has N nodes, where N is the number of atoms present in the cluster, and the output layer has 1 node. Each hidden layer has a sigmoid as the activation function and L2-regularization. \\

{\bf Calculation of spin-phonon coupling coefficients.}
All the relaxation rates presented in the main manuscript have been obtained using the spin-phonon coupling coefficients calculated from the machine learning model which was trained with $D$ values obtained from DFT calculations. The numerical differentiation method has been used to calculate the spin-phonon coupling coefficients. In the numerical differentiation method, one-dimensional grids of 6 points and two-dimensional grids of 36 points have been used to determine the first-order and second-order coefficients of spin-phonon coupling. Machine learning models have been used to calculate the $D$ values for all the points. Each one of these 2-dimensional grids of points was interpolated using a third-order polynomial, $P(x,y) = a_{30}x^3 +a_{03}y^3 +a_{21}x^2y+a_{12}xy^2 +a_{20}x^2 +a_{02}y^2 + a_{11}xy + a_{10}x + a_{01}y + a_{00}.$ The coefficients $a_{01}$ and $a_{10}$ correspond to first-order spin-phonon coupling coefficients, $(\partial \hat{H}_{s}/\partial X_{i})$, and coefficient corresponding to $a_{11}$, $a_{20}$ and $a_{02}$ represents the mixed and pure second-order Cartesian derivatives $(\partial^{2} \hat{H}_{s}/\partial X_{i}\partial X_{j})$, where the indexes $i,j$ runs over all the 3N degrees of freedom of the cell. Once these parameters have been computed, the spin-phonon coupling coefficients are obtained by means of the relations
\begin{equation}
    \left( \frac{\partial \hat{H}_{s}}{\partial q_{{\alpha}}} \right )= \sum_{i}^{3N} \sqrt{\frac{\hbar}{\omega_{\alpha}m_{i}}} L_{i\alpha}\left( \frac{\partial \hat{H}_{s}}{\partial X_i} \right )\:,
    \label{num_diff}
\end{equation}
and
\begin{equation}
    \left( \frac{\partial^2 \hat{H}_{s}}{\partial q_{{\alpha}}\partial q_{{\beta}}} \right )= \sum_{ij}^{3N} \sqrt{\frac{\hbar}{\omega_{\alpha}m_{i}}} \sqrt{\frac{\hbar}{\omega_{\beta}m_{j}}} L_{i\alpha}L_{j\beta}\left( \frac{\partial^2 \hat{H}_{s}}{\partial X_i\partial X_j} \right )\:,
    \label{num_diff2}
\end{equation}
where $L_{i\alpha}$ are the eigenvectors of the Hessian matrix.\\

\vspace{0.2cm}
\noindent
\textbf{Supplementary Materials}\\
The full expression of the spin-phonon transition rates. Tables S1-S2. Figures S1-S10. 

\vspace{0.2cm}
\noindent
\textbf{Acknowledgements and Funding}\\
This project has received funding from the European Research Council (ERC) under the European Union’s Horizon 2020 research and innovation programme (grant agreement No. [948493]). Computational resources were provided by the Trinity College Research IT and the Irish Centre for High-End Computing (ICHEC).

\vspace{0.2cm}
\noindent
\textbf{Authors Contributions}\\
A.L. conceived the project. S.M. performed all simulations and analyzed the data. All authors contributed to the discussion of the results and to writing the manuscript.

\vspace{0.2cm}
\noindent
\textbf{Data Availability}\\
All the spin dynamics simulations discussed in the present manuscript have been carried out with the software MolForge v1.0.0\cite{lunghi2022toward,MolForge}. The latest version of MolForge is available at \textit{github.com/LunghiGroup/MolForge}. All data needed to evaluate the conclusions in the paper are present in the paper and/or the Supplementary Materials. Additional data related to this paper may be requested from the authors.

\vspace{0.2cm}
\noindent
\textbf{Conflict of interests}\\
The authors declare no competing interests.

%\bibliographystyle{naturemag}
%\bibliography{biblio,Ale,manual_bib}

\begin{thebibliography}{10}
\expandafter\ifx\csname url\endcsname\relax
  \def\url#1{\texttt{#1}}\fi
\expandafter\ifx\csname urlprefix\endcsname\relax\def\urlprefix{URL }\fi
\providecommand{\bibinfo}[2]{#2}
\providecommand{\eprint}[2][]{\url{#2}}

\bibitem{alkahtani2018fluorescent}
\bibinfo{author}{Alkahtani, M.~H.} \emph{et~al.}
\newblock \bibinfo{title}{Fluorescent nanodiamonds: past, present, and future}.
\newblock \emph{\bibinfo{journal}{Nanophotonics}} \textbf{\bibinfo{volume}{7}},
  \bibinfo{pages}{1423--1453} (\bibinfo{year}{2018}).

\bibitem{castelletto2020silicon}
\bibinfo{author}{Castelletto, S.} \& \bibinfo{author}{Boretti, A.}
\newblock \bibinfo{title}{Silicon carbide color centers for quantum
  applications}.
\newblock \emph{\bibinfo{journal}{Journal of Physics: Photonics}}
  \textbf{\bibinfo{volume}{2}}, \bibinfo{pages}{022001} (\bibinfo{year}{2020}).

\bibitem{balasubramanian2009ultralong}
\bibinfo{author}{Balasubramanian, G.} \emph{et~al.}
\newblock \bibinfo{title}{Ultralong spin coherence time in isotopically
  engineered diamond}.
\newblock \emph{\bibinfo{journal}{Nature materials}}
  \textbf{\bibinfo{volume}{8}}, \bibinfo{pages}{383--387}
  (\bibinfo{year}{2009}).

\bibitem{bar2013solid}
\bibinfo{author}{Bar-Gill, N.}, \bibinfo{author}{Pham, L.~M.},
  \bibinfo{author}{Jarmola, A.}, \bibinfo{author}{Budker, D.} \&
  \bibinfo{author}{Walsworth, R.~L.}
\newblock \bibinfo{title}{Solid-state electronic spin coherence time
  approaching one second}.
\newblock \emph{\bibinfo{journal}{Nature communications}}
  \textbf{\bibinfo{volume}{4}}, \bibinfo{pages}{1--6} (\bibinfo{year}{2013}).

\bibitem{schirhagl2014nitrogen}
\bibinfo{author}{Schirhagl, R.}, \bibinfo{author}{Chang, K.},
  \bibinfo{author}{Loretz, M.} \& \bibinfo{author}{Degen, C.~L.}
\newblock \bibinfo{title}{Nitrogen-vacancy centers in diamond: nanoscale
  sensors for physics and biology}.
\newblock \emph{\bibinfo{journal}{Annual review of physical chemistry}}
  \textbf{\bibinfo{volume}{65}}, \bibinfo{pages}{83--105}
  (\bibinfo{year}{2014}).

\bibitem{gruber1997scanning}
\bibinfo{author}{Gruber, A.} \emph{et~al.}
\newblock \bibinfo{title}{Scanning confocal optical microscopy and magnetic
  resonance on single defect centers}.
\newblock \emph{\bibinfo{journal}{Science}} \textbf{\bibinfo{volume}{276}},
  \bibinfo{pages}{2012--2014} (\bibinfo{year}{1997}).

\bibitem{jelezko2004observation}
\bibinfo{author}{Jelezko, F.}, \bibinfo{author}{Gaebel, T.},
  \bibinfo{author}{Popa, I.}, \bibinfo{author}{Gruber, A.} \&
  \bibinfo{author}{Wrachtrup, J.}
\newblock \bibinfo{title}{Observation of coherent oscillations in a single
  electron spin}.
\newblock \emph{\bibinfo{journal}{Physical review letters}}
  \textbf{\bibinfo{volume}{92}}, \bibinfo{pages}{076401}
  (\bibinfo{year}{2004}).

\bibitem{grotz2011sensing}
\bibinfo{author}{Grotz, B.} \emph{et~al.}
\newblock \bibinfo{title}{Sensing external spins with nitrogen-vacancy
  diamond}.
\newblock \emph{\bibinfo{journal}{New Journal of Physics}}
  \textbf{\bibinfo{volume}{13}}, \bibinfo{pages}{055004}
  (\bibinfo{year}{2011}).

\bibitem{tetienne2013spin}
\bibinfo{author}{Tetienne, J.-P.} \emph{et~al.}
\newblock \bibinfo{title}{Spin relaxometry of single nitrogen-vacancy defects
  in diamond nanocrystals for magnetic noise sensing}.
\newblock \emph{\bibinfo{journal}{Physical Review B}}
  \textbf{\bibinfo{volume}{87}}, \bibinfo{pages}{235436}
  (\bibinfo{year}{2013}).

\bibitem{togan2010quantum}
\bibinfo{author}{Togan, E.} \emph{et~al.}
\newblock \bibinfo{title}{Quantum entanglement between an optical photon and a
  solid-state spin qubit}.
\newblock \emph{\bibinfo{journal}{Nature}} \textbf{\bibinfo{volume}{466}},
  \bibinfo{pages}{730--734} (\bibinfo{year}{2010}).

\bibitem{wolfowicz2021quantum}
\bibinfo{author}{Wolfowicz, G.} \emph{et~al.}
\newblock \bibinfo{title}{Quantum guidelines for solid-state spin defects}.
\newblock \emph{\bibinfo{journal}{Nature Reviews Materials}}
  \textbf{\bibinfo{volume}{6}}, \bibinfo{pages}{906--925}
  (\bibinfo{year}{2021}).

\bibitem{gottscholl2020initialization}
\bibinfo{author}{Gottscholl, A.} \emph{et~al.}
\newblock \bibinfo{title}{Initialization and read-out of intrinsic spin defects
  in a van der waals crystal at room temperature}.
\newblock \emph{\bibinfo{journal}{Nature materials}}
  \textbf{\bibinfo{volume}{19}}, \bibinfo{pages}{540--545}
  (\bibinfo{year}{2020}).

\bibitem{gottscholl2021room}
\bibinfo{author}{Gottscholl, A.} \emph{et~al.}
\newblock \bibinfo{title}{Room temperature coherent control of spin defects in
  hexagonal boron nitride}.
\newblock \emph{\bibinfo{journal}{Science Advances}}
  \textbf{\bibinfo{volume}{7}}, \bibinfo{pages}{eabf3630}
  (\bibinfo{year}{2021}).

\bibitem{stern2022room}
\bibinfo{author}{Stern, H.~L.} \emph{et~al.}
\newblock \bibinfo{title}{Room-temperature optically detected magnetic
  resonance of single defects in hexagonal boron nitride}.
\newblock \emph{\bibinfo{journal}{Nature communications}}
  \textbf{\bibinfo{volume}{13}}, \bibinfo{pages}{618} (\bibinfo{year}{2022}).

\bibitem{walker19685}
\bibinfo{author}{Walker, M.}
\newblock \bibinfo{title}{At 5 spin--lattice relaxation rate for non-kramers
  ions}.
\newblock \emph{\bibinfo{journal}{Canadian Journal of Physics}}
  \textbf{\bibinfo{volume}{46}}, \bibinfo{pages}{1347--1353}
  (\bibinfo{year}{1968}).

\bibitem{jarmola2012temperature}
\bibinfo{author}{Jarmola, A.}, \bibinfo{author}{Acosta, V.},
  \bibinfo{author}{Jensen, K.}, \bibinfo{author}{Chemerisov, S.} \&
  \bibinfo{author}{Budker, D.}
\newblock \bibinfo{title}{Temperature-and magnetic-field-dependent longitudinal
  spin relaxation in nitrogen-vacancy ensembles in diamond}.
\newblock \emph{\bibinfo{journal}{Physical review letters}}
  \textbf{\bibinfo{volume}{108}}, \bibinfo{pages}{197601}
  (\bibinfo{year}{2012}).

\bibitem{degen2017quantum}
\bibinfo{author}{Degen, C.~L.}, \bibinfo{author}{Reinhard, F.} \&
  \bibinfo{author}{Cappellaro, P.}
\newblock \bibinfo{title}{Quantum sensing}.
\newblock \emph{\bibinfo{journal}{Reviews of modern physics}}
  \textbf{\bibinfo{volume}{89}}, \bibinfo{pages}{035002}
  (\bibinfo{year}{2017}).

\bibitem{lunghi2019phonons}
\bibinfo{author}{Lunghi, A.} \& \bibinfo{author}{Sanvito, S.}
\newblock \bibinfo{title}{How do phonons relax molecular spins?}
\newblock \emph{\bibinfo{journal}{Science advances}}
  \textbf{\bibinfo{volume}{5}}, \bibinfo{pages}{eaax7163}
  (\bibinfo{year}{2019}).

\bibitem{escalera2019exploring}
\bibinfo{author}{Escalera-Moreno, L.}, \bibinfo{author}{Baldov{\'\i}, J.~J.},
  \bibinfo{author}{Gaita-Ari{\~n}o, A.} \& \bibinfo{author}{Coronado, E.}
\newblock \bibinfo{title}{Exploring the high-temperature frontier in molecular
  nanomagnets: from lanthanides to actinides}.
\newblock \emph{\bibinfo{journal}{Inorganic Chemistry}}
  \textbf{\bibinfo{volume}{58}}, \bibinfo{pages}{11883--11892}
  (\bibinfo{year}{2019}).

\bibitem{lunghi2020limit}
\bibinfo{author}{Lunghi, A.} \& \bibinfo{author}{Sanvito, S.}
\newblock \bibinfo{title}{The limit of spin lifetime in solid-state electronic
  spins}.
\newblock \emph{\bibinfo{journal}{The Journal of Physical Chemistry Letters}}
  \textbf{\bibinfo{volume}{11}}, \bibinfo{pages}{6273--6278}
  (\bibinfo{year}{2020}).

\bibitem{reta2021ab}
\bibinfo{author}{Reta, D.}, \bibinfo{author}{Kragskow, J.~G.} \&
  \bibinfo{author}{Chilton, N.~F.}
\newblock \bibinfo{title}{Ab initio prediction of high-temperature magnetic
  relaxation rates in single-molecule magnets}.
\newblock \emph{\bibinfo{journal}{Journal of the American Chemical Society}}
  \textbf{\bibinfo{volume}{143}}, \bibinfo{pages}{5943--5950}
  (\bibinfo{year}{2021}).

\bibitem{lunghi2022toward}
\bibinfo{author}{Lunghi, A.}
\newblock \bibinfo{title}{Toward exact predictions of spin-phonon relaxation
  times: An ab initio implementation of open quantum systems theory}.
\newblock \emph{\bibinfo{journal}{Science Advances}}
  \textbf{\bibinfo{volume}{8}}, \bibinfo{pages}{eabn7880}
  (\bibinfo{year}{2022}).

\bibitem{mondal2022unraveling}
\bibinfo{author}{Mondal, S.} \& \bibinfo{author}{Lunghi, A.}
\newblock \bibinfo{title}{Unraveling the contributions to spin--lattice
  relaxation in kramers single-molecule magnets}.
\newblock \emph{\bibinfo{journal}{Journal of the American Chemical Society}}
  (\bibinfo{year}{2022}).

\bibitem{astner2018solid}
\bibinfo{author}{Astner, T.} \emph{et~al.}
\newblock \bibinfo{title}{Solid-state electron spin lifetime limited by
  phononic vacuum modes}.
\newblock \emph{\bibinfo{journal}{Nature materials}}
  \textbf{\bibinfo{volume}{17}}, \bibinfo{pages}{313--317}
  (\bibinfo{year}{2018}).

\bibitem{tang2022first}
\bibinfo{author}{Tang, H.}, \bibinfo{author}{Barr, A.~R.},
  \bibinfo{author}{Wang, G.}, \bibinfo{author}{Cappellaro, P.} \&
  \bibinfo{author}{Li, J.}
\newblock \bibinfo{title}{First-principles calculation of the
  temperature-dependent transition energies in spin defects}.
\newblock \emph{\bibinfo{journal}{arXiv preprint arXiv:2205.02791}}
  (\bibinfo{year}{2022}).

\bibitem{cambria2022temperature}
\bibinfo{author}{Cambria, M.} \emph{et~al.}
\newblock \bibinfo{title}{Temperature-dependent phonon-induced relaxation of
  the nitrogen-vacancy spin triplet in diamond}.
\newblock \emph{\bibinfo{journal}{arXiv preprint arXiv:2209.14446}}
  (\bibinfo{year}{2022}).

\bibitem{lunghi2022spin}
\bibinfo{author}{Lunghi, A.}
\newblock \bibinfo{title}{Spin-phonon relaxation in magnetic molecules: Theory,
  predictions and insights}.
\newblock \emph{\bibinfo{journal}{arXiv preprint arXiv:2202.03776}}
  (\bibinfo{year}{2022}).

\bibitem{bayliss2020optically}
\bibinfo{author}{Bayliss, S.} \emph{et~al.}
\newblock \bibinfo{title}{Optically addressable molecular spins for quantum
  information processing}.
\newblock \emph{\bibinfo{journal}{Science}} \textbf{\bibinfo{volume}{370}},
  \bibinfo{pages}{1309--1312} (\bibinfo{year}{2020}).

\bibitem{garlatti2022critical}
\bibinfo{author}{Garlatti, E.} \emph{et~al.}
\newblock \bibinfo{title}{The critical role of ultra-low energy vibrations in
  the relaxation dynamics of molecular qubits}.
\newblock \emph{\bibinfo{journal}{Nature Communications}} \bibinfo{pages}{In
  Press} (\bibinfo{year}{2023}).

\bibitem{garlatti2020unveiling}
\bibinfo{author}{Garlatti, E.} \emph{et~al.}
\newblock \bibinfo{title}{Unveiling phonons in a molecular qubit with
  four-dimensional inelastic neutron scattering and density functional theory}.
\newblock \emph{\bibinfo{journal}{Nature communications}}
  \textbf{\bibinfo{volume}{11}}, \bibinfo{pages}{1--10} (\bibinfo{year}{2020}).

\bibitem{gugler2018ab}
\bibinfo{author}{Gugler, J.} \emph{et~al.}
\newblock \bibinfo{title}{Ab initio calculation of the spin lattice relaxation
  time t 1 for nitrogen-vacancy centers in diamond}.
\newblock \emph{\bibinfo{journal}{Physical Review B}}
  \textbf{\bibinfo{volume}{98}}, \bibinfo{pages}{214442}
  (\bibinfo{year}{2018}).

\bibitem{MolForge}
\bibinfo{author}{Lunghi, A.}
\newblock \bibinfo{title}{Molforge v.1.0.0}
  \bibinfo{pages}{DOI:10.5281/zenodo.7596042} (\bibinfo{year}{2023}).

\bibitem{yao2006theory}
\bibinfo{author}{Yao, W.}, \bibinfo{author}{Liu, R.-B.} \&
  \bibinfo{author}{Sham, L.}
\newblock \bibinfo{title}{Theory of electron spin decoherence by interacting
  nuclear spins in a quantum dot}.
\newblock \emph{\bibinfo{journal}{Physical Review B}}
  \textbf{\bibinfo{volume}{74}}, \bibinfo{pages}{195301}
  (\bibinfo{year}{2006}).

\bibitem{witzel2006quantum}
\bibinfo{author}{Witzel, W.} \& \bibinfo{author}{Sarma, S.~D.}
\newblock \bibinfo{title}{Quantum theory for electron spin decoherence induced
  by nuclear spin dynamics in semiconductor quantum computer architectures:
  Spectral diffusion of localized electron spins in the nuclear solid-state
  environment}.
\newblock \emph{\bibinfo{journal}{Physical Review B}}
  \textbf{\bibinfo{volume}{74}}, \bibinfo{pages}{035322}
  (\bibinfo{year}{2006}).

\bibitem{seo2016quantum}
\bibinfo{author}{Seo, H.} \emph{et~al.}
\newblock \bibinfo{title}{Quantum decoherence dynamics of divacancy spins in
  silicon carbide}.
\newblock \emph{\bibinfo{journal}{Nature communications}}
  \textbf{\bibinfo{volume}{7}}, \bibinfo{pages}{1--9} (\bibinfo{year}{2016}).

\bibitem{van1940paramagnetic}
\bibinfo{author}{Van~Vleck, J.}
\newblock \bibinfo{title}{Paramagnetic relaxation times for titanium and chrome
  alum}.
\newblock \emph{\bibinfo{journal}{Physical Review}}
  \textbf{\bibinfo{volume}{57}}, \bibinfo{pages}{426} (\bibinfo{year}{1940}).

\bibitem{shrivastava1983theory}
\bibinfo{author}{Shrivastava, K.}
\newblock \bibinfo{title}{Theory of spin--lattice relaxation}.
\newblock \emph{\bibinfo{journal}{physica status solidi (b)}}
  \textbf{\bibinfo{volume}{117}}, \bibinfo{pages}{437--458}
  (\bibinfo{year}{1983}).

\bibitem{simin2017locking}
\bibinfo{author}{Simin, D.} \emph{et~al.}
\newblock \bibinfo{title}{Locking of electron spin coherence above 20 ms in
  natural silicon carbide}.
\newblock \emph{\bibinfo{journal}{Physical Review B}}
  \textbf{\bibinfo{volume}{95}}, \bibinfo{pages}{161201}
  (\bibinfo{year}{2017}).

\bibitem{kanai2022generalized}
\bibinfo{author}{Kanai, S.} \emph{et~al.}
\newblock \bibinfo{title}{Generalized scaling of spin qubit coherence in over
  12,000 host materials}.
\newblock \emph{\bibinfo{journal}{Proceedings of the National Academy of
  Sciences}} \textbf{\bibinfo{volume}{119}}, \bibinfo{pages}{e2121808119}
  (\bibinfo{year}{2022}).

\bibitem{naydenov2011dynamical}
\bibinfo{author}{Naydenov, B.} \emph{et~al.}
\newblock \bibinfo{title}{Dynamical decoupling of a single-electron spin at
  room temperature}.
\newblock \emph{\bibinfo{journal}{Physical Review B}}
  \textbf{\bibinfo{volume}{83}}, \bibinfo{pages}{081201}
  (\bibinfo{year}{2011}).

\bibitem{de2020temperature}
\bibinfo{author}{de~Guillebon, T.}, \bibinfo{author}{Vindolet, B.},
  \bibinfo{author}{Roch, J.-F.}, \bibinfo{author}{Jacques, V.} \&
  \bibinfo{author}{Rondin, L.}
\newblock \bibinfo{title}{Temperature dependence of the longitudinal spin
  relaxation time t 1 of single nitrogen-vacancy centers in nanodiamonds}.
\newblock \emph{\bibinfo{journal}{Physical Review B}}
  \textbf{\bibinfo{volume}{102}}, \bibinfo{pages}{165427}
  (\bibinfo{year}{2020}).

\bibitem{acosta2010temperature}
\bibinfo{author}{Acosta, V.~M.} \emph{et~al.}
\newblock \bibinfo{title}{Temperature dependence of the nitrogen-vacancy
  magnetic resonance in diamond}.
\newblock \emph{\bibinfo{journal}{Physical review letters}}
  \textbf{\bibinfo{volume}{104}}, \bibinfo{pages}{070801}
  (\bibinfo{year}{2010}).

\bibitem{thiering2018theory}
\bibinfo{author}{Thiering, G.} \& \bibinfo{author}{Gali, A.}
\newblock \bibinfo{title}{Theory of the optical spin-polarization loop of the
  nitrogen-vacancy center in diamond}.
\newblock \emph{\bibinfo{journal}{Physical Review B}}
  \textbf{\bibinfo{volume}{98}}, \bibinfo{pages}{085207}
  (\bibinfo{year}{2018}).

\bibitem{albrecht2013coupling}
\bibinfo{author}{Albrecht, A.}, \bibinfo{author}{Retzker, A.},
  \bibinfo{author}{Jelezko, F.} \& \bibinfo{author}{Plenio, M.~B.}
\newblock \bibinfo{title}{Coupling of nitrogen vacancy centres in nanodiamonds
  by means of phonons}.
\newblock \emph{\bibinfo{journal}{New Journal of Physics}}
  \textbf{\bibinfo{volume}{15}}, \bibinfo{pages}{083014}
  (\bibinfo{year}{2013}).

\bibitem{golter2016optomechanical}
\bibinfo{author}{Golter, D.~A.}, \bibinfo{author}{Oo, T.},
  \bibinfo{author}{Amezcua, M.}, \bibinfo{author}{Stewart, K.~A.} \&
  \bibinfo{author}{Wang, H.}
\newblock \bibinfo{title}{Optomechanical quantum control of a nitrogen-vacancy
  center in diamond}.
\newblock \emph{\bibinfo{journal}{Physical review letters}}
  \textbf{\bibinfo{volume}{116}}, \bibinfo{pages}{143602}
  (\bibinfo{year}{2016}).

\bibitem{whiteley2019spin}
\bibinfo{author}{Whiteley, S.~J.} \emph{et~al.}
\newblock \bibinfo{title}{Spin--phonon interactions in silicon carbide
  addressed by gaussian acoustics}.
\newblock \emph{\bibinfo{journal}{Nature Physics}}
  \textbf{\bibinfo{volume}{15}}, \bibinfo{pages}{490--495}
  (\bibinfo{year}{2019}).

\bibitem{kuhne2020cp2k}
\bibinfo{author}{K{\"u}hne, T.~D.} \emph{et~al.}
\newblock \bibinfo{title}{Cp2k: An electronic structure and molecular dynamics
  software package-quickstep: Efficient and accurate electronic structure
  calculations}.
\newblock \emph{\bibinfo{journal}{The Journal of Chemical Physics}}
  \textbf{\bibinfo{volume}{152}}, \bibinfo{pages}{194103}
  (\bibinfo{year}{2020}).

\bibitem{goedecker1996separable}
\bibinfo{author}{Goedecker, S.}, \bibinfo{author}{Teter, M.} \&
  \bibinfo{author}{Hutter, J.}
\newblock \bibinfo{title}{Separable dual-space gaussian pseudopotentials}.
\newblock \emph{\bibinfo{journal}{Physical Review B}}
  \textbf{\bibinfo{volume}{54}}, \bibinfo{pages}{1703} (\bibinfo{year}{1996}).

\bibitem{perdew1996generalized}
\bibinfo{author}{Perdew, J.~P.}, \bibinfo{author}{Burke, K.} \&
  \bibinfo{author}{Ernzerhof, M.}
\newblock \bibinfo{title}{Generalized gradient approximation made simple}.
\newblock \emph{\bibinfo{journal}{Physical review letters}}
  \textbf{\bibinfo{volume}{77}}, \bibinfo{pages}{3865} (\bibinfo{year}{1996}).

\bibitem{grimme2010consistent}
\bibinfo{author}{Grimme, S.}, \bibinfo{author}{Antony, J.},
  \bibinfo{author}{Ehrlich, S.} \& \bibinfo{author}{Krieg, H.}
\newblock \bibinfo{title}{A consistent and accurate ab initio parametrization
  of density functional dispersion correction (dft-d) for the 94 elements
  h-pu}.
\newblock \emph{\bibinfo{journal}{The Journal of chemical physics}}
  \textbf{\bibinfo{volume}{132}}, \bibinfo{pages}{154104}
  (\bibinfo{year}{2010}).

\bibitem{neese2020orca}
\bibinfo{author}{Neese, F.}, \bibinfo{author}{Wennmohs, F.},
  \bibinfo{author}{Becker, U.} \& \bibinfo{author}{Riplinger, C.}
\newblock \bibinfo{title}{The orca quantum chemistry program package}.
\newblock \emph{\bibinfo{journal}{The Journal of chemical physics}}
  \textbf{\bibinfo{volume}{152}}, \bibinfo{pages}{224108}
  (\bibinfo{year}{2020}).

\bibitem{becke1993new}
\bibinfo{author}{Becke, A.~D.}
\newblock \bibinfo{title}{A new mixing of hartree--fock and local
  density-functional theories}.
\newblock \emph{\bibinfo{journal}{The Journal of chemical physics}}
  \textbf{\bibinfo{volume}{98}}, \bibinfo{pages}{1372--1377}
  (\bibinfo{year}{1993}).

\bibitem{adamo1999toward}
\bibinfo{author}{Adamo, C.} \& \bibinfo{author}{Barone, V.}
\newblock \bibinfo{title}{Toward reliable density functional methods without
  adjustable parameters: The pbe0 model}.
\newblock \emph{\bibinfo{journal}{The Journal of chemical physics}}
  \textbf{\bibinfo{volume}{110}}, \bibinfo{pages}{6158--6170}
  (\bibinfo{year}{1999}).

\end{thebibliography}

\end{document}